\documentclass[preprint,superscriptaddress,prb,aps,amsmath,amssymb]{revtex4-1}

\usepackage{graphicx}
\usepackage{amsmath}
\usepackage{amssymb}

\begin{document}

\title{Group-velocity slowdown in quantum-dots and quantum-dot molecules}

\author{Stephan Michael}
\affiliation{Department of Physics and Research Center OPTIMAS, University of Kaiserslautern, P.O. Box 3049, 67653 Kaiserslautern, Germany}
\author{ Weng W. Chow}
\affiliation{Semiconductor Materials and Device Sciences Department, Sandia
  National Laboratories, Albuquerque, NM 87185-1086, USA}
\author{Hans Christian Schneider}
\affiliation{Department of Physics and Research Center OPTIMAS, University of Kaiserslautern, P.O. Box 3049, 67653 Kaiserslautern, Germany}

\begin{abstract}
We investigate theoretically the slowdown of optical pulses due to quantum-coherence effects in
InGaAs-based quantum dots and quantum dot molecules.  Simple models  for the electronic structure of quantum dots and, in particular, quantum-dot molecules are described and calibrated using numerical simulations. It is shown how these models can be used to design optimized quantum-dot molecules for quantum coherence applications. The wave functions and energies obtained from the optimizations are used as input for a microscopic calculation of the quantum-dot material
dynamics including carrier scattering and polarization dephasing. The achievable group velocity slowdown in quantum-coherence $V$ schemes consisting of quantum-dot molecule states  is shown to be substantially higher than
what is achievable from similar transitions in typical InGaAs-based single
quantum dots.
\end{abstract}

\maketitle

\section{Introduction}

Quantum coherence effects arise from interference in the transition amplitudes between quantum states in the presence of a coherent light field.~\cite{Fleischhauer,intro4,scully1,intro2,intro3,mompart} 
Perhaps the best known of these effects are electromagnetically induced transparency (EIT) or slow light.  For many years, there have been proposals to realize quantum coherence effects
in few-level systems in solids~\cite{intro10,intro7,yang1,sarkar1,chuang-prb04:exc-pop-pulsation,boyd-pra04:coupled-resonator,Nikonov1,intro5} and, in particular, in
semiconductors.~\cite{intro8,intro9,intro11,hau1,intro6,intro13,intro10,yang1,sarkar1}
Slow light has been
achieved in semiconductor quantum wells (QWs) with coherent
population oscillations of excitons.~\cite{chuang-prb04:exc-pop-pulsation,Palinginis-APL} Other approaches inlcude slow light in photonic crystals.~\cite{Kondo}

Semiconductor QDs exhibit electron and hole states with discrete energies, and are reminiscent of atomic few-level systems.~\cite{chang1,intro12,qcpinsqd}. The dephasing of the quantum coherences, however, is much different 
from atomic systems. In particular, electron-hole transitions in semiconductors typically have short dephasing times which are detrimental for quantum coherence effects and limit the achievable group velocity slowdown, even in QDs,~\cite{jmo,apl:quantum-coherence,mork_JOSAB,NielsenProp}. Depending on
the levels that are connected by drive and probe fields,  $\Lambda$, $V$ and
ladder schemes can be realized.~\cite{Fleischhauer} A direct  comparison of these different setups in the framework of an
 atomic-like model with dephasing constants points toward the $V$ scheme as being the most useful and optimizable setup for group-velocity reduction.~\cite{mork_JOSAB}

The reason for investigating  $V$ type schemes instead of the
 $\Lambda$ schemes treated in earlier papers of us~\cite{qcpinsqd,apl:quantum-coherence,jmo} is that in $\Lambda$ scheme the quantum coherence connects two hole states. The hole states in GaAs-based semiconductor QDs are generally closely spaced and 
the electron-phonon interaction with polaronic broadening
efficiently couples them and leads to a pronounced dephasing for coherences involving hole states.
Because the drive or probe (electron-hole)
polarization is susceptible to the same dominant hole contributions of the
dephasing, the dephasing of the probe and quantum coherence are roughly of equal
size, which is not a good condition for quantum coherence effects. In this case, no group-velocity slowdown can be achieved with a CW drive pulse. Instead, a short drive pulse is necessary to slow down the probe
pulse,~\cite{apl:quantum-coherence,jmo} but the time window during which the
probe pulse is slowed down, is quite short.~\cite{dissertation} 

The present paper analyzes $V$-type quantum coherence schemes in the framework of a microscopic model, both for QDs and QD molecules.
It is a companion to our recent paper~\cite{qdm_prb}, in which we showed theoretically that one can achieve slowdown of optical pulses 
in InGaAs-based  QD molecules that is much larger than in $V$-type coherence schemes for single QDs systems. As explained in Ref.~\onlinecite{qdm_prb}, both the single QD and QD molecule $V$ schemes perform better than the $\Lambda$ schemes analyzed in our earlier studies~\cite{jmo,apl:quantum-coherence}, because the dephasing rates for the probe and quantum coherence are significantly different in $\Lambda$ and $V$ schemes. The encouraging results for the achievable group velocity slowdown in QD molecules contained in Ref.~\onlinecite{qdm_prb} are based on an optimized design for the QD molecule states that leads to a long lived coherence between two electronic levels used in the quantum coherence schemes.
  
In this paper, we give more details about the ``optimization'' of the QD and QD molecule structures used as input for the microscopic calculation of scattering and dephasing contributions for  polarization (and level population). A detailed description of the scattering and dephasing contributions in quantum coherence $V$ schemes is contained in Ref.~\onlinecite{qdm_prb}, here we focus on the modeling of the electronic structure of QDs and QD molecules to maximize the achievable slowdown. We describe comparatively simple models for the QD states, which can be calibrated by more realistic numerical calculations, but also permit us to vary important properties of the QD states by changing one (or a few) meaningful parameters, such as the depth of the QD confinement potential and/or the distance between the QDs making up the QD molecule. For the optimization of the QD structures, the figure of merit is a quasi-equilibrium slow-down factor determined by its group-velocity
reduction in the frequency domain. The slow down factor is calculated using a microscopic many-particle approach including  carrier scattering and polarization dephasing. 

The paper is organized as follows.
In Section~\ref{SBE} we give a brief review over the theory of semiconductor material
dynamics, i.e. the semiconductor Bloch equations. Section~\ref{SL_II} is devoted to single quantum dots, and investigates the influence of the
confinement potential and the lattice temperature on achievable slowdown for
InGaAs single QDs. Section~\ref{CH_QDM} is concerned with QD molecules. Here, we describe our model for QD molecules and show in some detail how to compute the electronic energies and states,
and highlight the features of our optimized QD molecule for
group-velocity slowdown. The results for group-velocity slowdown are compared with those of single QDs.

\section{Semiconductor Bloch Equations \label{SBE}}

In this section, we review the theory of semiconductor material dynamics,
which is necessary to describe the $V$-scheme of the QD system.
First of all, we introduce the optical field written in the form
\begin{equation}
\vec{E}(t) =\frac{1}{2}\hat{x}[ \mathcal{E}(t)e^{-i\omega t}
    +\mathcal{E}(t) e^{i\omega t}] \label{E1}
\end{equation}
where $\hat{x}$ is the polarization unit vector in $x$ direction, 
and $\omega$ is the frequency of the field $\vec{E}$. The corresponding macroscopic
polarization has the form
\begin{equation}
\vec{P}(t) =\frac{1}{2}\hat{x}[ \mathcal{P}(t) e^{-i\omega t}
    +\mathcal{P}^{*}(t)e^{i\omega t}]  \label{P1}
\end{equation}
where $\mathcal{P}$ is the complex slowly varying envelope.
The macroscopic polarization $P$ is connected with the microscopic
polarization by
\begin{equation}
P=\frac{N_{d}}{L}\sum_{\alpha ,\beta }\mu _{\alpha\beta }p_{\alpha \beta } + c.c.
\label{eq_makro}
\end{equation}
where $N_{d}$ is the in-plane density of the QDs, $L$ is
the thickness of the region, in which the QD layer is
embedded, $\mu_{\alpha\beta }$ are the dipole matrix elements and
the summation index $\alpha$ or $\beta$ refers to QD system
electron or hole states, respectively.

The dynamics of the polarizations and carrier distributions at the
single-particle level are calculated in the framework of the
semiconductor Bloch equations for the reduced single-particle density
matrix. We denote in the following electron and hole levels in the QD $\alpha$ and~$\beta$, respectively.
For the $V$ system of interest in this paper one obtains the following
equations of motion for the ``interband'' polarizations,
$p_{\alpha \beta}$, and the ``intra(electron-)band'' polarizations~$p_{\alpha'\alpha''}$
\begin{align}
\begin{split}
\frac{\partial}{\partial t}p_{\beta\alpha } =&  - i \omega_{\alpha \beta }p_{\beta \alpha } - i\Omega _{\alpha \beta }\left(n_{\alpha }^{c}-n_{\beta }^{v}\right) 
  -i\sum_{\alpha'\neq \alpha}\Omega _{\alpha'\beta}p_{\alpha'\alpha}+S_{\beta \alpha}
\label{p-ab}
\end{split}\\
\begin{split}
\frac{\partial }{\partial t}p_{\alpha'\alpha''}
=& - i\omega _{\alpha''\alpha'}p_{\alpha'\alpha''}-i\Omega_{\alpha''\alpha'}\left( n_{\alpha''}^{c}-n_{\alpha'}^{c}\right) 
   +i\sum_{\beta'}\left( \Omega_{\alpha''\beta'}p_{\alpha'\beta'}-\Omega_{\beta'\alpha'}p_{\beta'\alpha''}\right) +S_{\alpha'\alpha''}
\label{p-aa}
\end{split}
\end{align}
In particular, the polarization $p_{e_{0} e_{1}}$ here is the \emph{quantum coherence}.
For the time evolution of the conduction and valence band populations,
$n_{\alpha }^{c}$ and $n_{\beta }^{v}$, one obtains
\begin{align}
\frac{\partial}{\partial t}n_{\alpha}^{c} &=i\sum_{\beta'}
\left( \Omega_{\alpha \beta'}p_{\alpha\beta'}
-\Omega_{\beta'\alpha}p_{\beta'\alpha}\right) +S_{\alpha \alpha } \label{n-a} \\
\frac{\partial }{\partial t}n_{\beta }^{v}& =i\sum_{\alpha'}
\left( \Omega _{\beta \alpha'}p_{\beta\alpha'}-\Omega_{\alpha'\beta}p_{\alpha'\beta}\right) +
S_{\beta \beta }
\label{n-b}
\end{align}
The coherent contributions of the above equations contains transition
frequencies $\omega_{\alpha \beta}$ and renormalized Rabi
frequencies $\Omega_{\alpha \beta}=\hbar^{-1} \mu_{\alpha \beta} E \left(
t \right) + \Omega_{\alpha \beta}^{\text{HF}}$ with $E \left( t
\right) = \frac{\mathcal{E}\left(t\right)}{2} [e^{-i\omega t}+e^{i\omega t}]$ are renormalized by
excitation-dependent Hartree-Fock (HF) contributions resulting from the
Coulomb interaction, as discussed, e.g., in
Refs.~\onlinecite{jmo,apl:quantum-coherence,qcpinsqd}. 
The correlation contributions are generally denoted by $S$ and contain the influence of carrier-carrier and carrier-phonon interactions beyond the Hartree-Fock level. In particular,
$S_{\alpha,\alpha}$ and $S_{\beta,\beta}$ describe scattering contributions in the dynamical equations for the electron and hole distributions as well as
dephasing $S_{\beta \alpha}$, $S_{\alpha'\alpha''}$ in the dynamical equations
for the coherences. The correlation contributions $S$ are derived and the
explicit equations are given in Ref.~\onlinecite{qdm_prb}.

For the calculation of the implied 
Coulomb matrix elements and carrier-phonon interaction matrix
elements in the dots-in-a-well system, QD and QW
states has to be considered.
Our approach can not handle the whole dots-in-a-well system in
one \textquotedblleft box\textquotedblright . Such an approach would
naturally yield localized and delocalized eigenfunctions that are orthogonal
to each other. We have to treat the calculation of the three-dimensional QD states separately
from the calculation of the QW states. 
To describe the combined system we orthogonalize the QW states to the QD
states as described in Ref.~\onlinecite{prb64:115315}.
The outcome of this are localized and delocalized eigenfunctions that are orthogonal
to each other as used in Ref.~\onlinecite{qdm_prb}.

To determine the spectral gain and group-velocity slowdown in a $V$-system, we solve the
dynamical equations~\eqref{p-ab}--\eqref{n-b} for a strong cw drive field with
fixed angular frequency $\omega_d$ and a weak cw probe field with angular
frequency $\omega_p$. From the steady-state value of the polarization
$\mathcal{P}$ we determine the gain via
\begin{equation}
g(\omega_{p}) =-\frac{\omega_{p}}{2\varepsilon_{0}cn_{b}\mathcal{E}_{p}}
\Im[\mathcal{P}]
\end{equation}
and refractive-index change
\begin{equation}
\delta n(\omega_{p}) =-\frac{1}{2\varepsilon_{0}
  n_{b}\mathcal{E}_{p}}\Re[\mathcal{P}]
\end{equation}
where $n_{b}$ is the background refractive index of the host
material. The group-velocity slowdown factor is defined by
$S\left( \omega _{p}\right) =n_{b}
+\omega_{p} \frac{d\left( \delta n\right) }{d\omega_{p}}\equiv n_b + S'(\omega_p)$, but we will consider only the contribution from the index change 
\begin{equation}
S'(\omega_p)=\omega_{p} \frac{d\left( \delta n\right) }{d\omega_{p}}
\end{equation}
in order to remove the static contribution, which describes the change in group velocity due to the background refractive index as compared to vacuum.
 
\section{Single quantum dots \label{SL_II}}

\begin{figure}[]
\centering
\includegraphics[trim=0cm 0cm 0cm 0cm,clip,scale=0.6,angle=0]{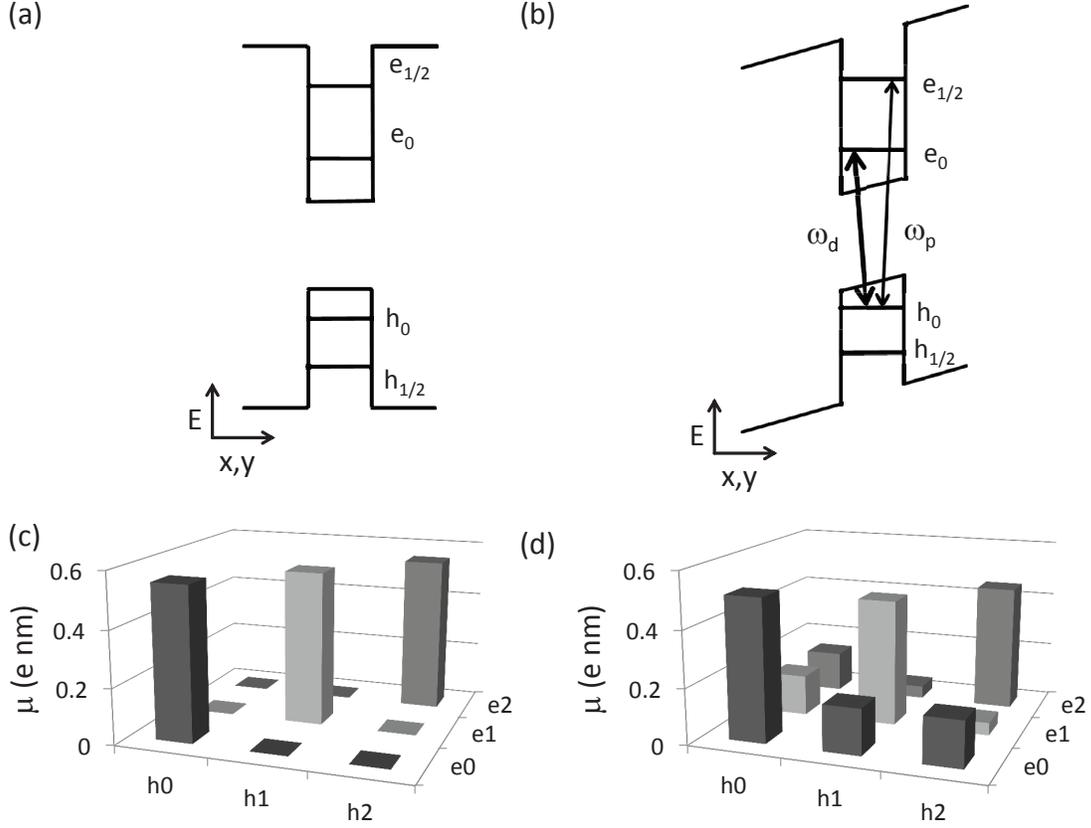}
\caption{Energy spectrum and setup of the cw drive and probe fields for
the slowing down of the probe field in a V-scheme with (a) and without (b)
static electric field. 
Dipole interband matrix elements $\mu$ for a deep QD with (c) and
  without (d) static electric field. The static field makes the V-scheme
  possible because it breaks the symmetry responsible for the vanishing of the
  off-diagonal dipole matrix elements.}
\label{fig_z1}
\end{figure}

In this section we calculate the group-velocity slowdown for single QDs with a $V$-type configuration of probe and drive pulses, see Figure~\ref{fig_z1}(b) for a sketch of the resulting band lineup. The results of this Section make use of a simplified QD model in order to show--including a microscopic calculation of the relevant dephasing of the quantum coherences-- the possibility of group velocity slowdown with a CW drive.  These results are also used as a baseline to measure the improvements for group-velocity reduction that come from using optimized QD molecules, which will be discussed in the next section.

We describe briefly the simple QD model used in this section to calculate the matrix elements needed for the calculation of the microscopic QD dynamics described in Section~\ref{SBE}. We assume that the QDs are contained in a surrounding quantum well, and that  the envelope function for electron and hole states can be written as a product of a wave function in the growth direction $z$ of the QW and an in-plane part
\begin{equation}
\Phi_{3\text{D}}(r,\varphi ,z)=N\, \Phi _{\parallel }(r,\varphi ) \Phi_{\perp }(z)
\label{eq:2D1D-wavefunction}
\end{equation}
where $N$ is a normalization constant. The in-plane confinement potential is assumed to be harmonic and is completely specified by the harmonic oscillator level spacing $\hbar\omega_{\text{HO}}$, which is chosen in accordance with measured/calculated values. The in-plane part of the wave function is then given by eigenfunctions of the two-dimensional harmonic oscillator, cf. Ref.~\cite{Hawrylak-book,prb64:115315} The band lineup of the QD in this simple model is therefore defined by the spacing of electron and hole levels, as well as the fundamental band-gap.

In the following we use the QD model described above to define two different QDs and
investigate the group-velocity slowdown performance of these QDs  by calculating the spectral gain and group-velocity slowdown from
the QD material dynamics described in section~\ref{SBE}. In particular, we compare the results of the
slowdown factor for different lattice temperatures and cw drive intensities.

\begin{table}[]
\begin{center}
\begin{tabular}{||c||c||}
\hline\hline
Shallow Dot & Deep Dot \\ \hline\hline
\begin{tabular}{lll}
& E$_{e}$ (meV) & E$_{h}$ (meV) \\ 
e/h$_{0}$ & \multicolumn{1}{c}{$-70$} & \multicolumn{1}{c}{$30$} \\ 
e/h$_{1/2}$ & \multicolumn{1}{c}{$-30$} & \multicolumn{1}{c}{$15$}%
\end{tabular}
& 
\begin{tabular}{lll}
& E$_{e}$ (meV) & E$_{h}$ (meV) \\ 
e/h$_{0}$ & \multicolumn{1}{c}{$-150$} & \multicolumn{1}{c}{$50$} \\ 
e/h$_{1/2}$ & \multicolumn{1}{c}{$-60$} & \multicolumn{1}{c}{$20$}%
\end{tabular}
\\ \hline\hline
\end{tabular}%
\end{center}
\caption{Electron (e) and hole (h) energies of single-particle states
  in the single QDs.}
\label{table-1}
\end{table}

The $V$-scheme employed here is shown schematically in figure \ref{fig_z1}. 
We assume an ensemble of InAs QDs
embedded in a GaAs QW with a width of $16$ nm. 
We will investigate a shallow and a deep QD with three confined electron
and hole states. Thus we have one doubly degenerate excited state and one
ground state with the energy values in Table~\ref{table-1}.
Using an analytical model only diagonal transitions are dipole allowed
because of symmetry considerations. However, to realize a $V$-scheme one
needs off-diagonal  interband
transitions. We achieve this by including a symmetry breaking static
electric field. To make off-diagonal dipole matrix elements appreciable, we
use an external electric field in the plane of the QW with a field
strength of $4.0$~mV~nm$^{-1}$ for the deep QD. The
diagonal and off-diagonal dipole moments with and without the external
electric field are shown in figure~\ref{fig_z1}. The dipole
matrix elements make a $V$-scheme with a drive-pulse between the electron
and hole ground state and a probe-pulse between the hole ground and the
excited electron states possible. The quantum coherence of the $V$ 
scheme is between the electron ground and the excited electron states. The
energy gap between the electron and hole ground state is taken to be around $1.2$~eV.

\begin{figure}[]
\centering
\includegraphics[trim=1cm 3cm 1cm 4cm,clip,scale=0.6,angle=0]{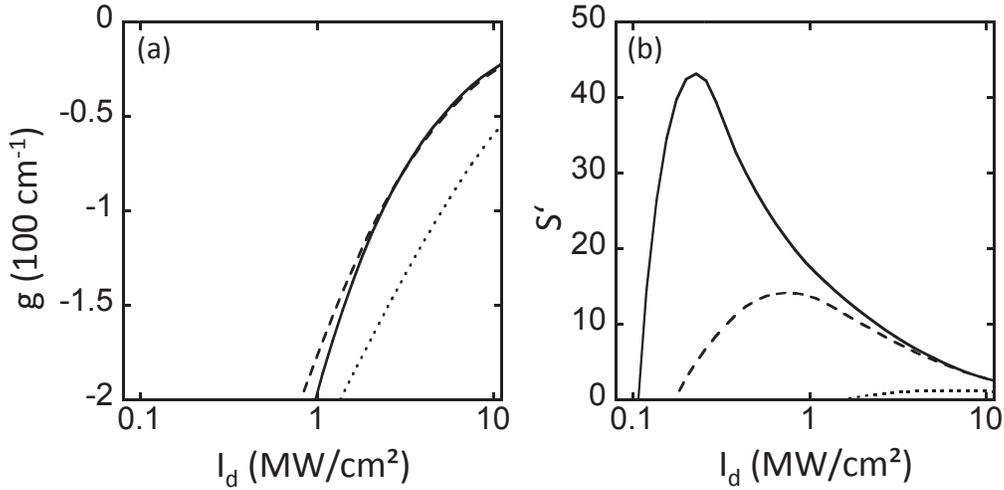}
\caption{Peak gain (a) and peak slowdown (b) versus drive intensity for a
  shallow (dotted line) and a deep (solid and dashed line) QD. The
lattice temperature is 150~K (solid line) and 300~K (dashed and dotted line).}
\label{fig3}
\end{figure}

We compare a shallow and a deep QD for slow light applications in
figure~\ref{fig3}. We choose a weak cw probe with intensity $45$~W/cm$^{2}$.
For a given cw drive
intensity the spectral gain and group-velocity slowdown can be determined 
as described in section~\ref{SBE} by running the calculation until 
a steady state for the probe polarization is reached.
Afterwards the peak gain and peak slowdown versus cw drive intensity 
can be plotted as shown in figure~\ref{fig3}.  

First, we discuss the shallow and deep QD results for 
a lattice temperature of $300$\,K.
Below a drive intensity of $0.1$\,MW/cm$^{2}$ we find a significant peak absorption
without peak slow-down for both QDs. Above a drive intensity of $0.1$\,MW/cm$^{2}$ the peak slow-down factor for similar peak
absorption values is higher for the deep QD. A optimum drive intensity
around $1.0$~MW/cm$^{2}$ facilitate only a small absorption and a
significant peak slowdown for the probe of the deep QD. For the shallow QD
no significant peak slowdown is reached. Therefore, a
maximum of slowdown with a minimum of absorption is accomplished better for the deep QD as for the
shallow QD. 
The results can be explained in the following way: If we want
to reach transparency with an appreciable slowdown, the dephasing rate of the
polarization from the cw probe has to significantly exceed the dephasing
rate of the quantum coherence. 
For the deep QD the energy spacing of the electron states is large  enough to
suppress the electronic-intersubband contribution of the carrier-phonon
dephasing, but the hole-intersubband contributions are still significant. 
Thus the carrier-phonon dephasing of the quantum coherence is small compared to the
carrier-phonon dephasing of the probe polarization.
But for the shallow QD the two carrier-phonon dephasing rates are of similar size. The
carrier-carrier dephasing rate of the quantum coherence is only
slightly different between the deep and the shallow QD. 
Therefore, compared to the shallow QD, the deep QD has a significantly 
smaller dephasing rate of the quantum coherence.

Figure \ref{fig3} shows also the results for the lattice temperature
dependence of peak gain and peak slowdown by comparing a lattice temperature of $150$\,K
with a lattice temperature of $300$\,K for the deep QD. 
Above a drive intensity of 
$0.1$\,MW/cm$^{2}$ the peak slowdown factor for
similar peak absorption values is higher for lower temperatures.
Thus, a slightly enhanced EIT with an improvement of slowdown is obtained for a
lattice temperature of $150$\,K compared to $300$\,K. The results can be
explained in the following way: Because the average phonon occupation is
reduced for lower temperatures, a smaller carrier-phonon dephasing rate
results for all polarizations. This reduction is more pronounced for the
interband and less pronounced for the quantum coherence, which is
already small. Additionally, the carrier-carrier dephasing rate exhibits a small, but not
significant, change for the quantum coherence between $300$\,K
and $150$\,K. 

Qualitatively, the picture that emerges for group velocity slowdown in single QDs based on a microscopic description of the quantum coherences is the following. The computed group-velocity slowdown of the $V$-type scheme is considerably larger than that achievable in a $\Lambda $-type scheme. In particular, one can obtain group-velocity slowdown with a CW drive field using a realistic model for the sources of dephasing of the quantum coherence. In a $\Lambda$ scheme, as investigated earlier by us, this is not possible and one needs to resort to pulsed drive fields, which lead to usable slowdown only in a short time window. In the $V$ configuration investigated here, a deep dot at low temperatures
gives the best results, but the carrier-carrier dephasing contributions, while already small,
cannot be significantly reduced and therefore set the  limit of
efficiency for slow light applications. In the next section we  show
how a reduction of the relevant carrier-phonon and especially
\emph{carrier-carrier} contributions to the dephasing can be realized by reducing the overlap of the
corresponding wave functions by using suitably designed QD \emph{molecules}.

\section{Quantum dot molecules\label{CH_QDM}}

In this section we first describe a semi-analytical model to investigate the electronic structure of QD
molecules. This approach allows us to specify  QD molecule ``designs'' by the sizes (and material composition) of the individual dots and the distance of
the single QDs. For these QD designs we can find, in an approximate way, the associated wave functions and energies. We show how the structural parameters of the QD molecules determine the energy levels and wave functions of the QD molecules. Finally, we choose QD molecule structure optimized for long-lived quantum coherences and calculate the group-velocity slowdown achievable in this structure.

Compared to the single QD model of the last section, we make an adjustment to the confinement potential because
an in-plane harmonic oscillator confinement potential has no finite size in plane. If one wants to combine two single dots QDs to a molecule, including the electronic coupling of the single QD states, it is necessary to 
determine the wave functions and energy levels of the single QDs from a \emph{finite} confinement potential. Since the details of the confinement potential do not decisively affect the final results, it is easier to work with a pillbox model for the single QDs,  instead of modifying the harmonic oscillator potential. Starting from the wave functions of this pillbox model for the single QDs, we make an ansatz for the QD molecule states similar to the linear combination of atomic
orbitals. As in the QD case, the Coulomb interaction between the electron or hole states is taken into account later in the dynamic calculation using the semiconductor Bloch equations. 

\subsection{Electronic structure of a cylindrical QD \label{app_dot}}

We follow here and in the next subsection closely the Appendix in Ref.~\onlinecite{qdm_prb}. For the Hamiltonian of the cylindrical QD in envelope approximation we use
\begin{equation}
H=-\frac{\hbar ^{2}}{2m}\nabla^2 +V(r,z)
\label{3DSchroedinger_cylinderQD}
\end{equation}
where the Laplacian~$\nabla^2$ and the confinement potential
\begin{equation}
V(r,z)=\left\{ 
\begin{array}{c}
0 \\  -V_{0}
\end{array}
\right. \left. 
\begin{array}{c}
\text{for }\left\vert z\right\vert >a \\ 
\text{for }\left\vert z\right\vert <a
\end{array}%
\right. \left. 
\begin{array}{c}
\text{or} \\ 
\text{and}%
\end{array}%
\right. \left. 
\begin{array}{c}
\left\vert r\right\vert >b \\ 
\left\vert r\right\vert <b%
\end{array}%
\right.
\label{3DPotential_cylinderQD}
\end{equation}%
are expressed in cylindrical coordinates. In this approximation, the QD ``design'' and, consequently, the electronic structure is fixed by the following parameters: the height of the QDs in $z$ 
direction $h=2a$, the diameter~$2b$, and the depth of
the confinement potential~$V_{0}$. The full envelope wave function $\Phi _{3D}$  is the solution of the
Schroedinger equation
\begin{equation}
H\Phi _{3\text{D}}=E\Phi _{3\text{D}}.
\end{equation}

Under the realistic assumption that the height is much smaller than the diameter of the QD, the electrons and holes are strongly localized in the growth
direction~$z$. If we assume the separability of the wave function for the
in-plane and the $z$~direction, the three dimensional Schroedinger equation
can be separated into a 2D and a 1D problem as in Eq.~(\ref{eq:2D1D-wavefunction}). Furthermore we assume, in $z$ direction,
\begin{equation}
\int \Phi_{\perp }^{*}(z) V(r,z) \Phi_{\perp }(z) \text{ d}z \approx -V_{0} \Theta(b-|r|) 
\end{equation}
and the corresponding approximation for the in-plane direction 
to obtain a self-consistent set of equations.

For the Schroedinger equation in $z$ direction we have
\begin{equation}
\left[ -\frac{\hbar ^{2}}{2m}\frac{\partial^{2}}{\partial z^{2}}
+\tilde{V}_{\perp }\,\Theta(a-|z|)\right] \Phi_{\perp }(z)=E_{\perp }\Phi _{\perp }(z)
\end{equation}
Here, $\tilde{V}_{\perp}=-V_0 + T_{\|}$ is an effective one-dimensional potential that contains
the in-plane kinetic energy $T_\|=\tilde{V}_\|-E_\|$ which, in turn, depends
on the in-plane eigenenergy~$E_\|$. 
Since these kinetic energies $T_{\|}$ are
not known, we use an iteration procedure to calculate the in-plane and $z$
eigenenergies. We start the iteration by setting $\tilde{V}_{\perp}$
equal to $-V_0$. 
We obtain for the symmetric eigenstates
\begin{equation}
\begin{split}
\Phi_{\perp ,n}^{S}(z)= B &\Theta ( |z| -a)
\cos(k_{n}a)e^{\kappa _{n}(a-|z|)}  
+ B\Theta (a-\left\vert z\right\vert )\cos(k_{n}z)
\end{split}
\end{equation}
and for the antisymmetric eigenstates
\begin{equation}
\begin{split}
\Phi_{\perp ,n}^{A}(z)=C&\Theta (|z| -a)
\text{sgn}\left(z\right) \sin(k_{n}a)e^{\kappa _{n}(a-| z|)}  
+ C\Theta (a-|z|)\sin(k_{n}z).
\end{split}
\end{equation}
Here $B$ and $C$ are normalization constants and we have defined
\begin{align}
\kappa_{n} &=\frac{\sqrt{2m|E_{\perp ,n}|}}{\hbar} \\
k_{n} &=\frac{\sqrt{2m\big( |\tilde{V}_{\perp }|-|E_{\perp ,n}|\big) }}{\hbar} 
\end{align}%
The eigenvalues $E_{\perp ,n}$ can be determined by the intersection $s_{n}=k_{n}a$ of the curves
\begin{align}
f(ka) &=\tan(ka) \\
g^{S}(ka) &=\frac{\sqrt{(k_{0}a)^{2}-(ka)^{2}}}{(ka)} 
\end{align}
or
\begin{equation}
g^{A}(ka)=\frac{-(ka)}{\sqrt{(k_{0}a)^{2}-(ka)^{2}}}
\end{equation}%
where $k_{0}=\sqrt{2m | \tilde{V}_{\perp}| / \hbar}$. Finally, the eigenvalues are
\begin{equation}
E_{\perp ,n}=\frac{\hbar ^{2}s_{n}^{2}}{2ma^{2}}- |\tilde{V}_{\perp}|.
\end{equation}

For the Schroedinger equation in the in-plane direction we have
\begin{equation}
\begin{split}
\Big[ -\frac{\hbar ^{2}}{2m}\big[ \frac{1}{r}\frac{\partial }{\partial r}
&\big( r\frac{\partial }{\partial r}\big) +\frac{1}{r^{2}}\frac{\partial
^{2}}{\partial \varphi ^{2}}\big] 
+ \tilde{V}_{\parallel}\Theta(b-r)\Big]
\Phi _{\parallel }(r,\varphi )=E_{\parallel }\Phi _{\parallel }(r,\varphi )
\end{split}
\label{2DSchroedinger_cylinderQD}
\end{equation}
The effective potential $\tilde{V}_{\parallel }=-V_0+T_{\perp}$
again includes a
contribution from the kinetic energy in growth-direction, which depends on the solution of the eigenvalue problem in $z$ direction, $T_\perp=\tilde{V}_{\perp}-E_{\perp}$. We start the iteration
by setting $\tilde{V}_{\parallel }$ equal to $-V_0$. Because of the symmetry of the potential
around the growth direction, the Hamiltonian commutes with the components of
the angular momentum operator ($\left[ H,l_{z}\right] =0$). Therefore the
two dimensional Schroedinger equation for the angular momentum projection
quantum number $m_{l}$ reduces to an effective one dimensional Schroedinger
equation. Resorting the terms we obtain
\begin{equation}
\begin{split}
\Big[ -\frac{\hbar ^{2}}{2m}\big( \frac{1}{r}\frac{\partial }{\partial r}
&+
\frac{\partial ^{2}}{\partial r^{2}}\big) 
+ \tilde{V}_\|\Theta(b-r)+\frac{\hbar ^{2}}{2m}\frac{m_{l}^{2}}{r^{2}}
\Big] \tilde{\Phi}_{\parallel }(r)  =E_{\parallel }\tilde{\Phi}(r)
\end{split}
\end{equation}
where
\begin{equation}
\Phi _{\parallel}(r,\varphi)=\frac{1}{\sqrt{2\pi }}e^{im\varphi }\tilde{\Phi}
_{\parallel ,m}(\varphi)
\end{equation}
This equation can be cast into the form of a Bessel differential equation. 
A solution of this differential equation inside the
QD is the Bessel function in $m$-th order of the first kind $J_{m}(kr)$, so that we have inside the QD
\begin{equation}
\widetilde{\Phi }_{\parallel ,m}(r) = A J_{m}(kr)
\end{equation}%
A solution outside the QD is the modified Bessel function $K_{m}(\kappa_{r}r)$. Outside the QD we therefore have
\begin{equation}
\widetilde{\Phi}_{\parallel ,m}\left( r\right) =BK_{m}(\kappa _{r}r).
\end{equation}
At $r=b$, $\Psi'_{\parallel }$ and $\Psi
_{\parallel }$ have to be continuous. With $k_{0}^{2}=\frac{2m}{\hbar^2}
( -\tilde{V}_{\parallel }) $ and $\kappa_{\mathrm{r}}=\sqrt{k_{0}^{2}-k^{2}}$,
the continuity condition yields
\begin{equation}
N(k)=\frac{J_{m}^{\prime }(kR)}{J_{m}(kR)}-\frac{K_{m}^{\prime }
(\sqrt{k_{0}^{2}-k^{2}}R)}{K_{m}(\sqrt{k_{0}^{2}-k^{2}}R)}=0
\end{equation}
All $k_{n}$ between $0$ and $k_{0}$ with $N(k)=0$ are allowed. For the
eigenvalues of the two dimensional problem we obtain
\begin{equation}
E_{\parallel ,n}=\frac{\hbar^2 \left( k_{n}^{2}-k_{0}^{2}\right) }{2m}
\end{equation}
In summary we have energy levels $E_{\parallel,n,m}$ and wave functions 
$\Phi_{\parallel ,n,m}$ with the quantum numbers $n$ and $m$. The states with
different $m$ and the same $n$ are degenerate.

For the approximate solution of the three dimensional problem we have
to solve the one- and two-dimensional eigenvalues in a self-consistent fashion by determining the updated potentials for the next iteration
step from the eigen-energies of the previous iteration. The procedure is quite efficient, and one obtains converged eigenvalues $E_{n_{z}n_{r}m}$ and wave functions $\Phi _{n_{z}n_{r}m}$ for the
pillbox-shaped QD after only a few iteration steps. The resulting energies and wave functions, obtained using optimized effective parameters, have been checked against
k$\cdot $p-calculations,~\cite{hackenbuchner,homepage} which include strain and piezoelectric effects.

\subsection{Electronic structure of a QD molecule \label{app_molecule}}

We assume that the QD molecules are stacked on top of each other, as can be achieved using vertically correlated
growth of QDs.~\cite{QDH2} With this method, QDs are grown
in layers on top of each other, separated by a spacer layer. 
For this type of QD molecules  we study  different QD heights and widths of the spacer layer
and analyze the resulting energy spectra and dipole moments. 

Using the electronic structure
for pillbox shaped QDs, we now couple these QDs to molecules. For this purpose we
introduce an ansatz similar to the linear combination of atomic orbitals.
We assume a QD
molecule consisting of two QDs, labeled $1$ and $2$. For QD $1$ and $2$ we
assume $N$ and $M$ bound states respectively. Further, for the uncoupled
QDs, we label the wave functions $\Phi _{1}^{n}$ and $\Phi _{2}^{m}$, the
eigenvalues $\varepsilon _{1}^{n}$ and $\varepsilon _{2}^{m}$ and the
potential $V_{a}$ and $V_{b}$, respectively. To determine the envelope wave
functions $\Phi $, and the corresponding eigenvalues $E$, of the electronically coupled
QDs we use a superposition of the following form 
\begin{equation}
\Phi =\sum_{n}c_{1}^{n}\Phi_{1}^{n}+\sum_{m}c_{2}^{m}\Phi_{2}^{m}
\label{QDM1}
\end{equation}
With the Hamiltonian 
\begin{equation}
\big( H_{0}+V_{a}+V_{b}\big) \Phi =E\Phi  
\label{QDM2}
\end{equation}%
we can apply a multiplication of $( \Phi _{1}^{j})^*$ and
a multiplication of $(\Phi _{2}^{k})^*$ respectively.
Therefore we obtain in matrix notation
\begin{equation}
\begin{pmatrix}
M_{1}^{jn} & M_{2}^{jm} \\ 
M_{3}^{kn} & M_{4}^{km}%
\end{pmatrix}%
\binom{c_{1}^{n}}{c_{2}^{m}}=%
\begin{pmatrix}
A_{1}^{jn} & A_{2}^{jm} \\ 
A_{3}^{kn} & A_{4}^{km}
\end{pmatrix}%
E\binom{c_{1}^{n}}{c_{2}^{m}}  \label{QDM3}
\end{equation}%
where%
\begin{align}
M_{1}^{jn} &=\varepsilon _{1}^{n}\delta _{jn}+\langle \Phi
_{1}^{j}\vert V_{b}\vert \Phi _{1}^{n}\rangle  \label{QDM4}
\\
M_{2}^{jm} &=\varepsilon _{2}^{m}\langle \Phi
_{1}^{j}\vert \Phi _{2}^{m}\rangle +\langle \Phi
_{1}^{j}\vert V_{a}\vert \Phi _{2}^{m}\rangle  \\
M_{3}^{kn} &=\varepsilon _{1}^{n}\langle  \Phi
_{2}^{k}\vert \Phi _{1}^{n}\rangle +\langle \Phi
_{2}^{k}\vert V_{b}\vert \Phi _{1}^{n}\rangle  \\
M_{4}^{km} &=\varepsilon _{2}^{m}\delta _{km}+\langle \Phi
_{2}^{k}\vert V_{a}\vert \Phi _{2}^{m}\rangle
\end{align}%
and
$A_{1}^{jn} =\delta _{jn}$, 
$A_{2}^{jm} =\langle \Phi _{1}^{j}\vert \Phi_{2}^{m}\rangle$,
$A_{3}^{kn} =\langle\Phi _{2}^{k}\vert \Phi_{1}^{n}\rangle$, as well as
$A_{4}^{km} =\delta _{km} $.
This generalized eigenvalue problem can be solved numerically with an eigenvalue-solver.~\cite{LARPACK} Because in this case matrix~$A$ is invertible,
its possible to reduce the generalized eigenvalue problem to an (ordinary)
eigenvalue problem. Therefore we have to solve
\begin{equation}
\left[ 
\begin{pmatrix}
A_{1}^{jn} & A_{2}^{jm} \\ 
A_{3}^{kn} & A_{4}^{km}%
\end{pmatrix}%
^{-1}%
\begin{pmatrix}
M_{1}^{jn} & M_{2}^{jm} \\ 
M_{3}^{kn} & M_{4}^{km}
\end{pmatrix}%
\right] \binom{c_{1}^{n}}{c_{2}^{m}}=E\binom{c_{1}^{n}}{c_{2}^{m}}
\label{QDM6}
\end{equation}
The eigenvalues and eigenfunctions of this equation have to be
understood as the single-particle result for the electronic structure of the
QD molecule, which can then be used as input in the many-particle
semiconductor Bloch equations.

Furthermore we want to consider an sufficiently weak external electric
field, i.e., an electric field that can be included in the LCAO calculation of the QD molecules.
For electrons, one includes in the potential $V_{a}+V_{b}$ in~\eqref{QDM2} a contribution from the electric field $Fz$ where $F$ is the electric field. For holes, the sign of the electric potential is reversed.
The results of this semi-analytical approach for QD molecules without electric
field were again checked against k$\cdot $p-calculation.~\cite{hackenbuchner,homepage} 
The approach was found to yield a qualitatively correct description of the electronic structure of the QD molecules studied in this paper.

\subsection{Examples of QD molecules\label{QDM_Examples}}

\begin{figure}[]
\centering
\includegraphics[trim=2cm 3cm 2cm 2cm,clip,scale=0.60,angle=0]{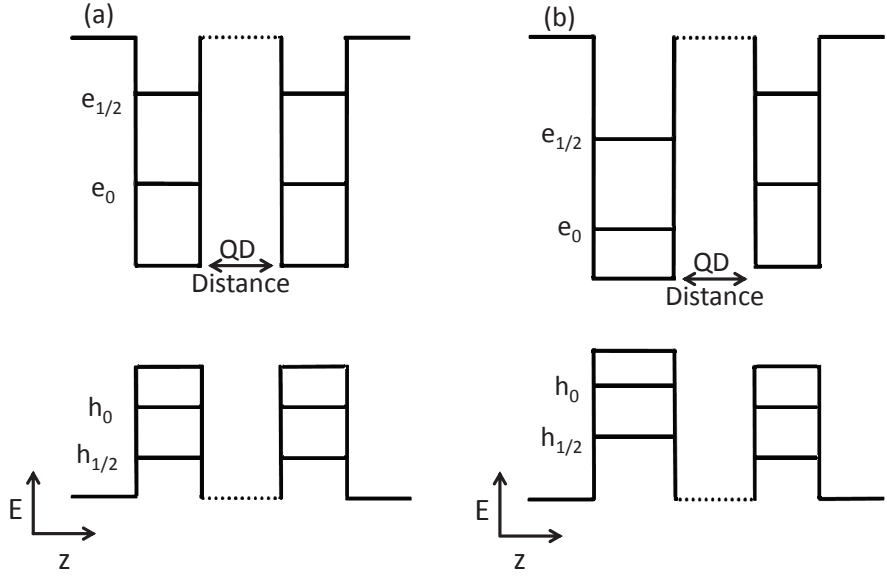}
\caption{Schematic picture of the geometry of the symmetric (a) and asymmetric
  (b) double QD molecules. The QD distance is varied as described in the text.}
\label{fig5}
\end{figure}

\begin{table}[]
\begin{center}
\begin{tabular}{||c||c||}
\hline\hline
Large Dot & Small Dot \\ \hline\hline
\begin{tabular}{lll}
& E$_{e}$ (meV) & E$_{h}$ (meV) \\ 
e/h$_{0}$ & \multicolumn{1}{c}{$-192.8$} & \multicolumn{1}{c}{$52.0$} \\ 
e/h$_{1/2}$ & \multicolumn{1}{c}{$-105.3$} & \multicolumn{1}{c}{$26.6$}%
\end{tabular}
& 
\begin{tabular}{lll}
& E$_{e}$ (meV) & E$_{h}$ (meV) \\ 
e/h$_{0}$ & \multicolumn{1}{c}{$-144.8$} & \multicolumn{1}{c}{$38.8$} \\ 
e/h$_{1/2}$ & \multicolumn{1}{c}{$-62.3$} & \multicolumn{1}{c}{$14.9$}%
\end{tabular}
\\ \hline\hline
\end{tabular}
\end{center}
\caption{Electron (e) and hole (h) energies of single-particle states
  in the single QDs forming the double QD molecules.}
\label{table-2}
\end{table}

As shown in figure~\ref{fig5} 
we assume two cylindrical QDs stacked in $z$-direction with a 
QD distance $d$ and an aligned in-plane center of the potential. 
We will compare the electronic structure and the dipole
matrix elements for molecules consisting of identical QDs,
e.g., two small QDs, and QDs of different sizes, i.e. a small and a large
QD. The parameters used in the semi-analytical model
are adjusted to sample QDs calculated using $k\cdot p$-theory. 
For the sample QD we choose an InAs cylindrical QD embedded in GaAs.
We assume for the cylindrical QD a diameter of $16$\,nm, 
a height of $3$\,nm, and call this one the large QD. 
The input material parameters are taken from Ref.~\onlinecite{material}.
The numerical reference calculation is done by using a single-band approximation for the electron states
and a $6\times 6$ k$\cdot $p-method for the hole states.~\cite{homepage} We find three
confined electron and three confined hole states. The energy gap between the
electron and the hole ground state of the QD comes out to be $1.2$\,eV, and the hole ground state is over 90\% heavy-hole like. Afterwards, we
adjust the parameters and calculate the large QD and the small QD using the semi-analytical
approach. For the small QD we assume a
diameter of $14$ nm and a height of $2.5$\,nm. 
The energy eigenvalues of both QDs for our adjusted semi-analytical model are given in table~\ref{table-2}.

\begin{figure}[]
\centering
\includegraphics[trim=2cm 0cm 2cm 0cm,clip,scale=0.3,angle=0]{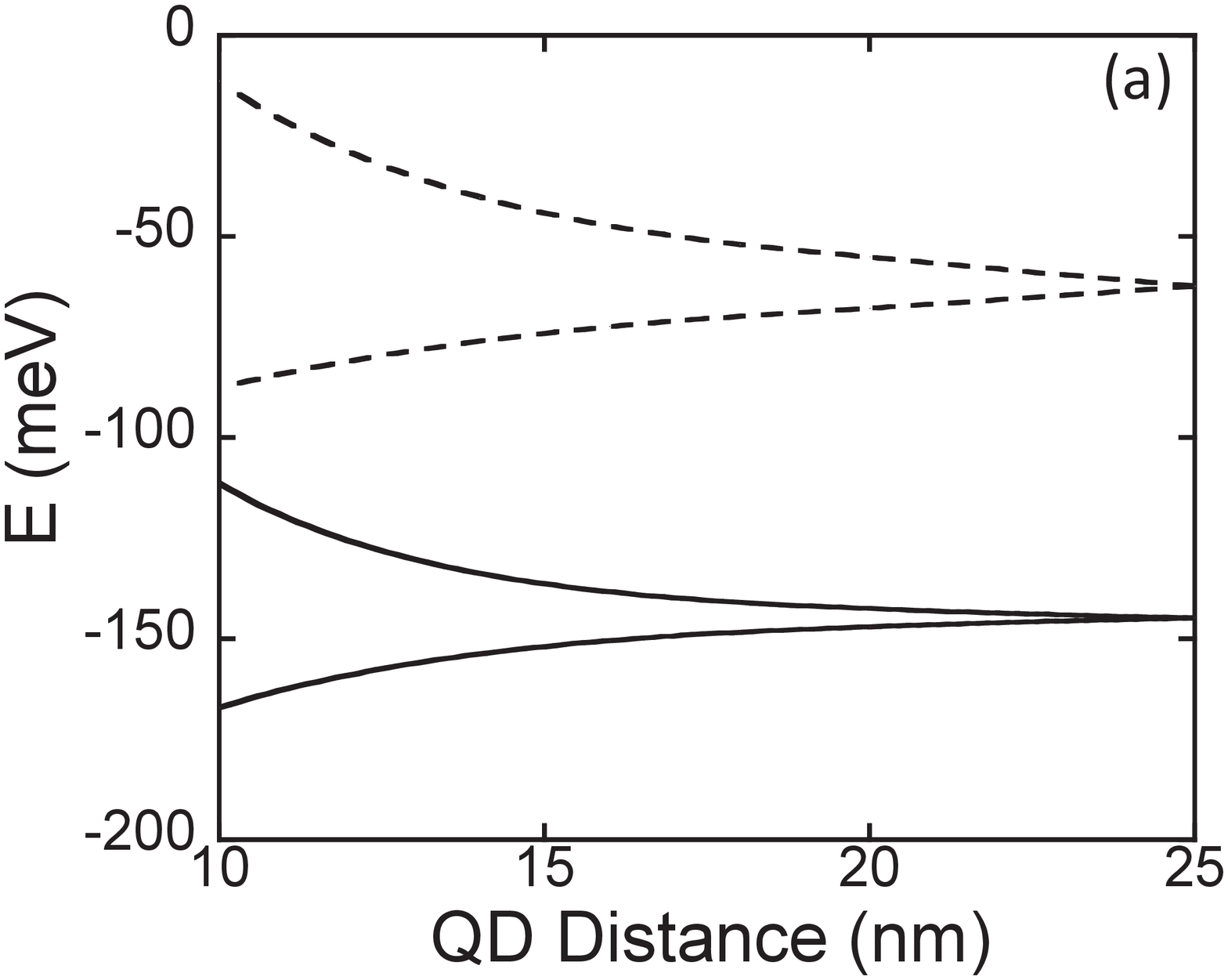}
\includegraphics[trim=2cm 0cm 2cm 0cm,clip,scale=0.3,angle=0]{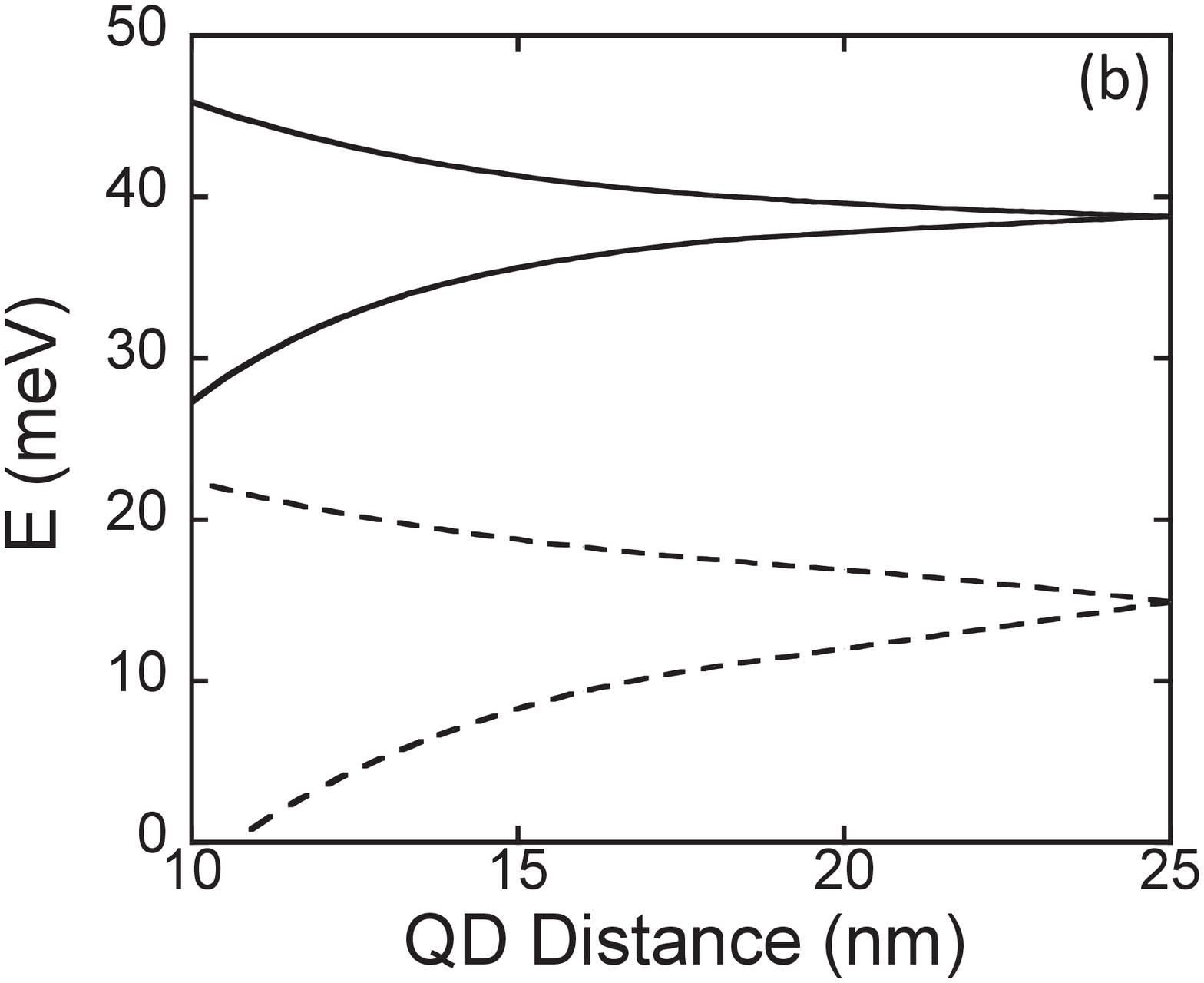}
\includegraphics[trim=2cm 0cm 2cm 0cm,clip,scale=0.3,angle=0]{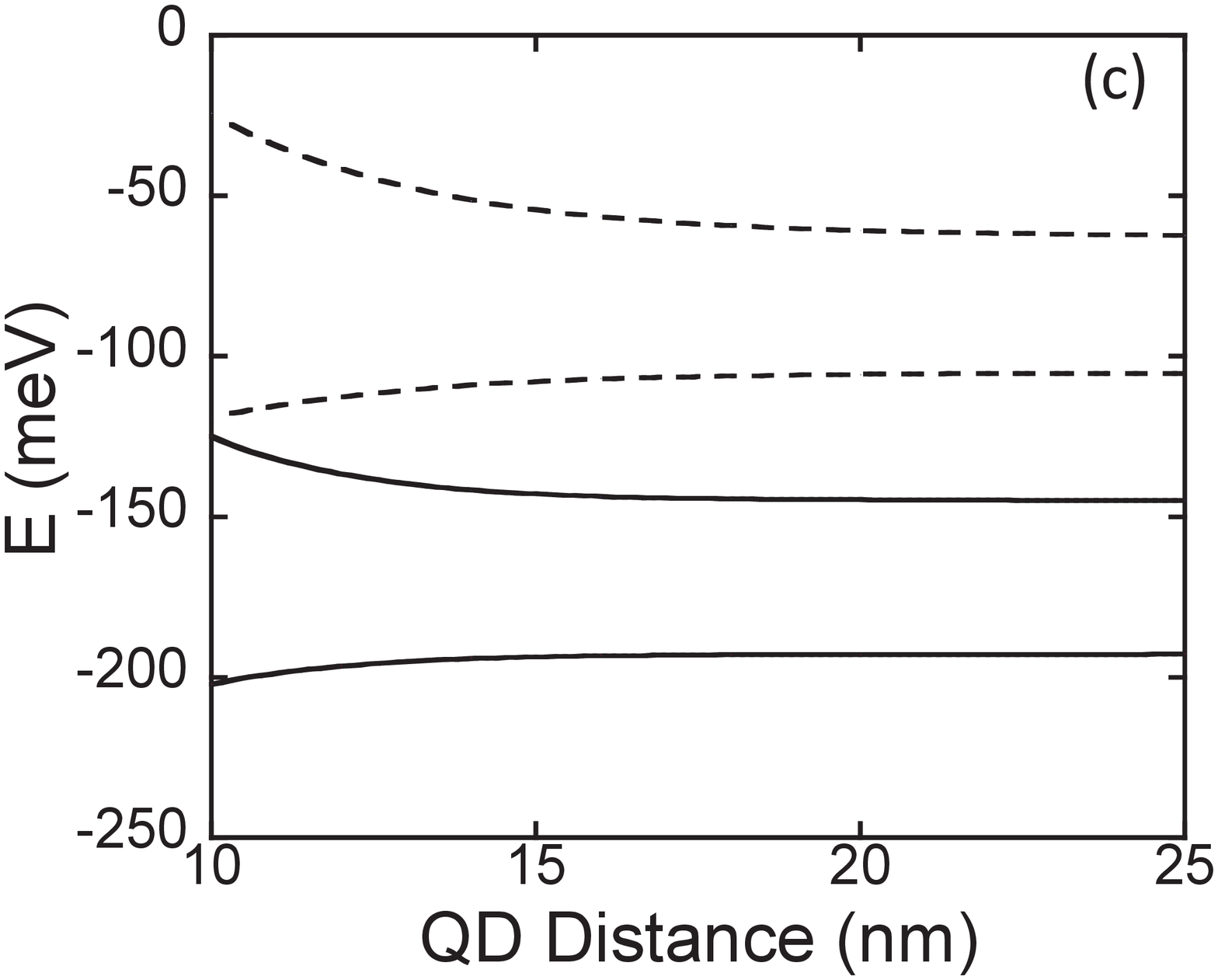}
\includegraphics[trim=2cm 0cm 2cm 0cm,clip,scale=0.3,angle=0]{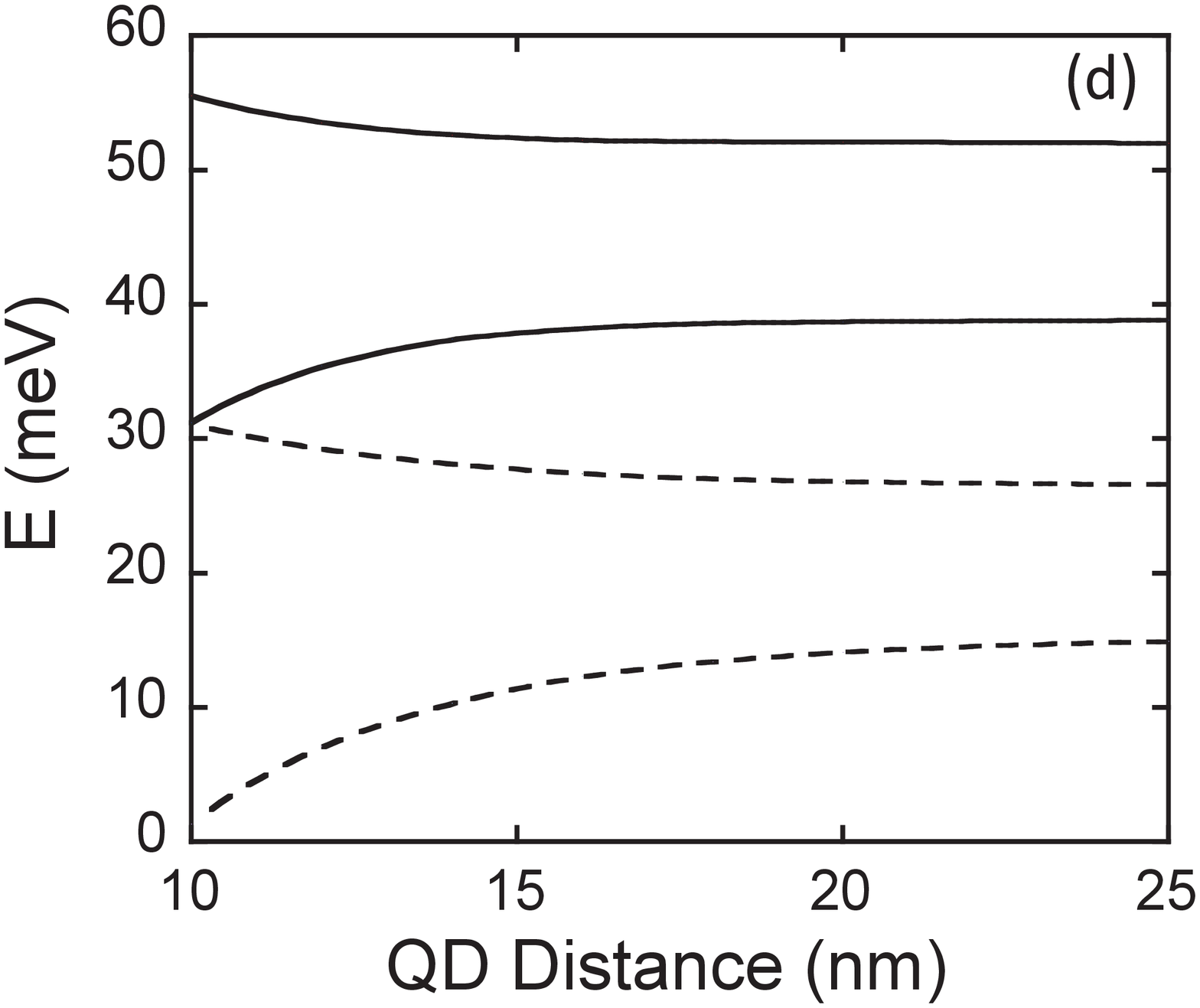}
\caption{Energy of the combined conduction (a,c) and valence band states (b,d)
 plotted over the QD distance between two identical (a,b) 
and two different sized QDs (c,d) forming a molecule.
The solid lines are the bonding and antibonding ground states and the dashed lines are the bonding and
antibonding degenerate first and second excited states.}
\label{qdm_energy1}
\end{figure}

\begin{figure}[]
\centering
\includegraphics[trim=0cm 7cm 0cm 3cm,clip,scale=0.60,angle=0]{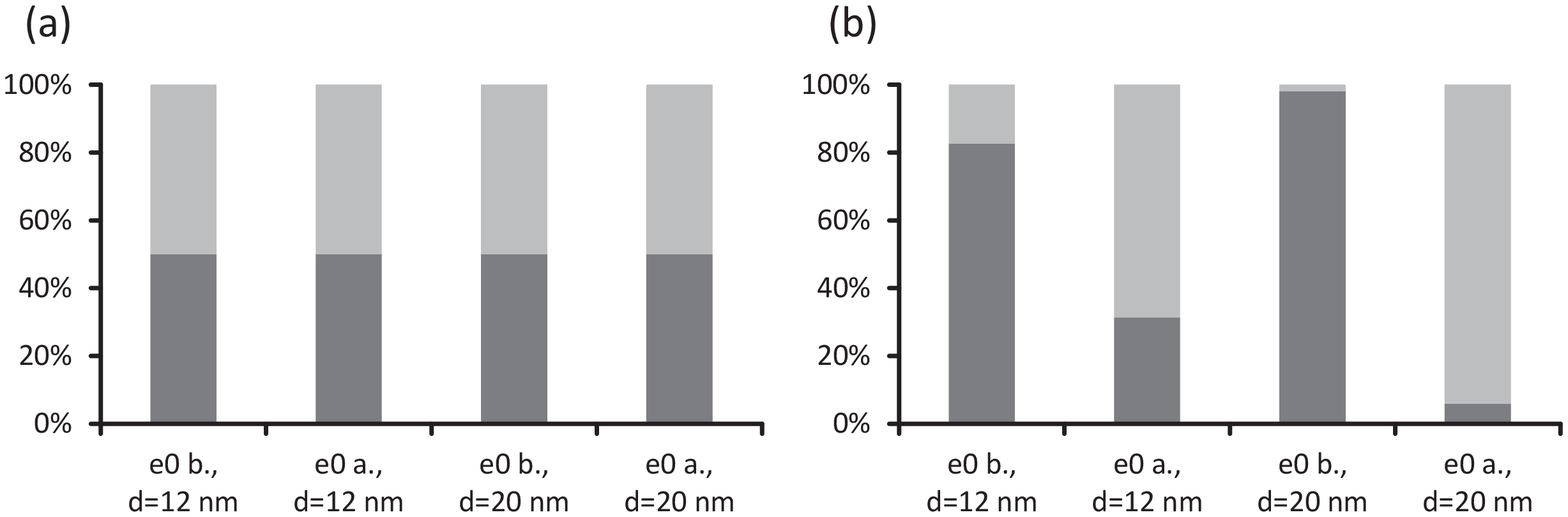}
\caption{Composition of the bonding and antibonding ground states for a distance of 12 nm and 20 nm between QDs of identical (a) and different (b) size.
The bottom QD is marked black and the top QD is marked grey. In (b) the bottom QD is the ``large'' QD and the top QD the ``small'' one.}
\label{qdm_energy2}
\end{figure}

First, we investigate a QD molecule consisting of 
two identical QDs and use the small QDs for the calculation. We start with a
QD distance of 
$25$\,nm between the QDs and repeat our calculation with smaller distances
until a QD distance
of $10$\,nm is reached. For each distance the wave functions of the QD molecule and
the corresponding energy values are determined. The electron and hole energy
values for these QD distances are plotted in figure~\ref{qdm_energy1}.
If the distance between the QDs is sufficiently large the bottom and the
top dot are electronically decoupled and can be considered as single QDs.
Therefore the lowest level and the higher levels of the bottom and top QDs are
degenerate. By reducing the distance, the interaction between the QDs becomes stronger, and a bonding and a antibonding state originating
from the degenerate states are obtained. As illustrated in figure~\ref{qdm_energy2}, for identical QDs the bonding
state $\Phi _{b}$ can be written as $\Phi _{b}=\frac{1}{\sqrt{2}}\left( \Phi_{bt}+\Phi _{t}\right) $ and the antibonding state $\Phi _{a}$ can be
written as $\Phi _{a}=\frac{1}{\sqrt{2}}\left( \Phi _{bt}-\Phi _{t}\right) $
over the whole range of possible distances, where $\Phi _{bt}$ is the state
originating from the bottom and $\Phi _{t}$ is the state originating from
the top QD.  When the
QDs are closer, the interaction becomes stronger and the energy separation
between the bonding and antibonding states is larger. For the energetically
lower lying states this energy separation is symmetric. For confined states
at higher energies the separation is somewhat suppressed for the antibonding
state. Qualitatively, the behavior of the bonding and antibonding states
over the considered distance range is similar for the conduction and the
valence band.

We now turn to QDs of different size. We couple a ``bottom'' large QD and a ``top'' small QD. Again we calculate the combined wave
functions and the corresponding eigenvalues for several QD distances between 25\,nm and 10\,nm, and plot the electron and hole energy
values in figure~\ref{qdm_energy1}.
If the distance between the QDs is sufficiently large
the bottom and the top dot are electronically decoupled and can be
considered as single QDs. By reducing the distance the QDs begin to
interact with each other and a bonding state originating from the large QD
and an antibonding state originating from the small QD is obtained (see
figure~\ref{qdm_energy1}). If the QDs are closer, the interaction
becomes stronger and the energy separation between the bonding and
antibonding states is enlarged. In contrast to the identical QDs the
composition of the bonding and antibonding states changes with the distance as
depicted in figure~\ref{qdm_energy2}.
For shorter distances the bonding states develop an increasing admixture from states
originating from the small QD and vice versa. 
Qualitatively,
the behavior of the bonding and antibonding states over the
range of distances considered here is again similar for the conduction and the valence band. Generally, comparing Figures~\ref{qdm_energy2} (a) and (b) shows that combining QDs of similar size to a molecule will lead to a more efficient mixing of states located at the individual QDs as compared to the combination of QDs with very different sizes.

\begin{figure}[]
\centering
\includegraphics[trim=1cm 0cm 1cm 0cm,clip,scale=0.3,angle=0]{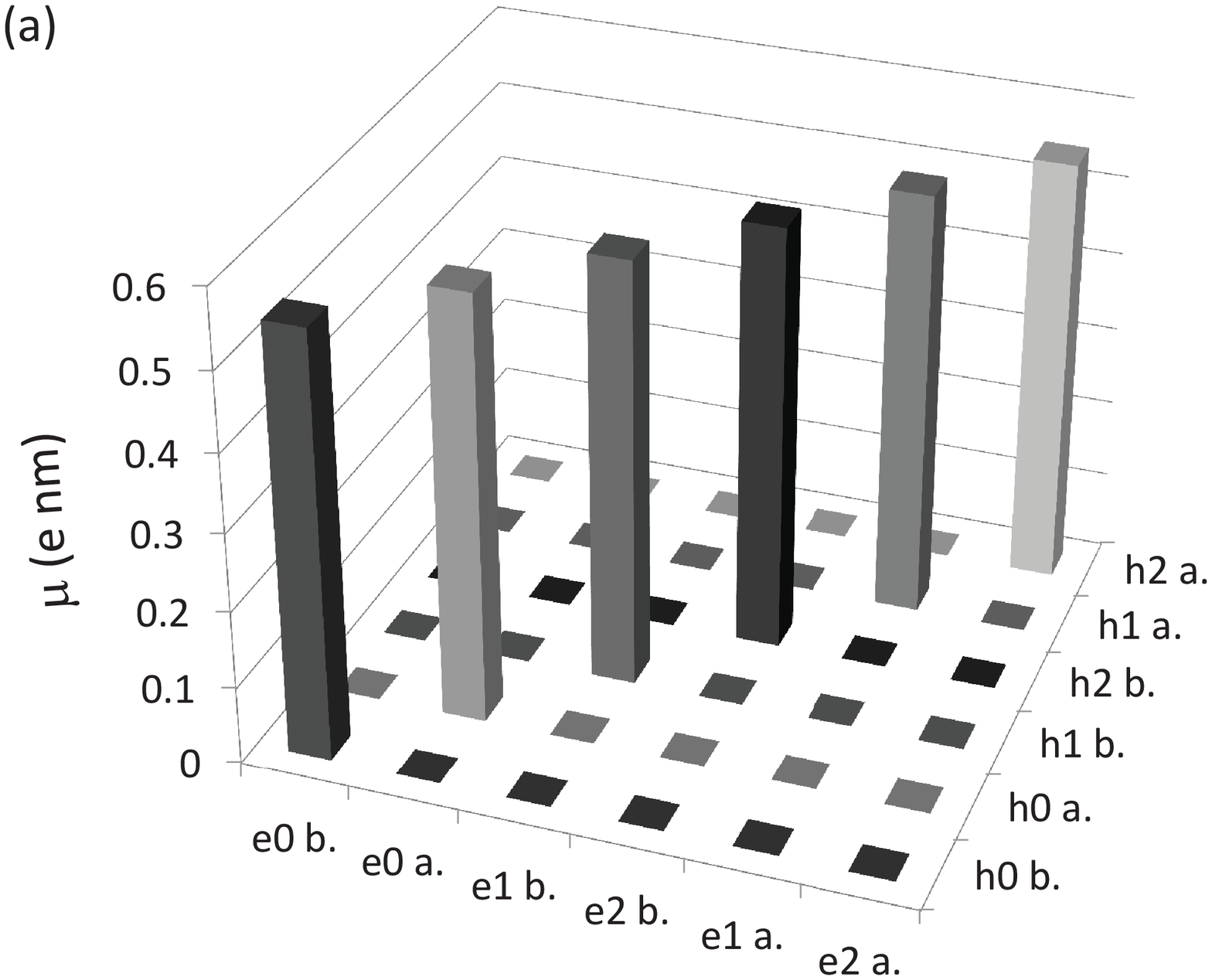}
\includegraphics[trim=1cm 0cm 1cm 0cm,clip,scale=0.3,angle=0]{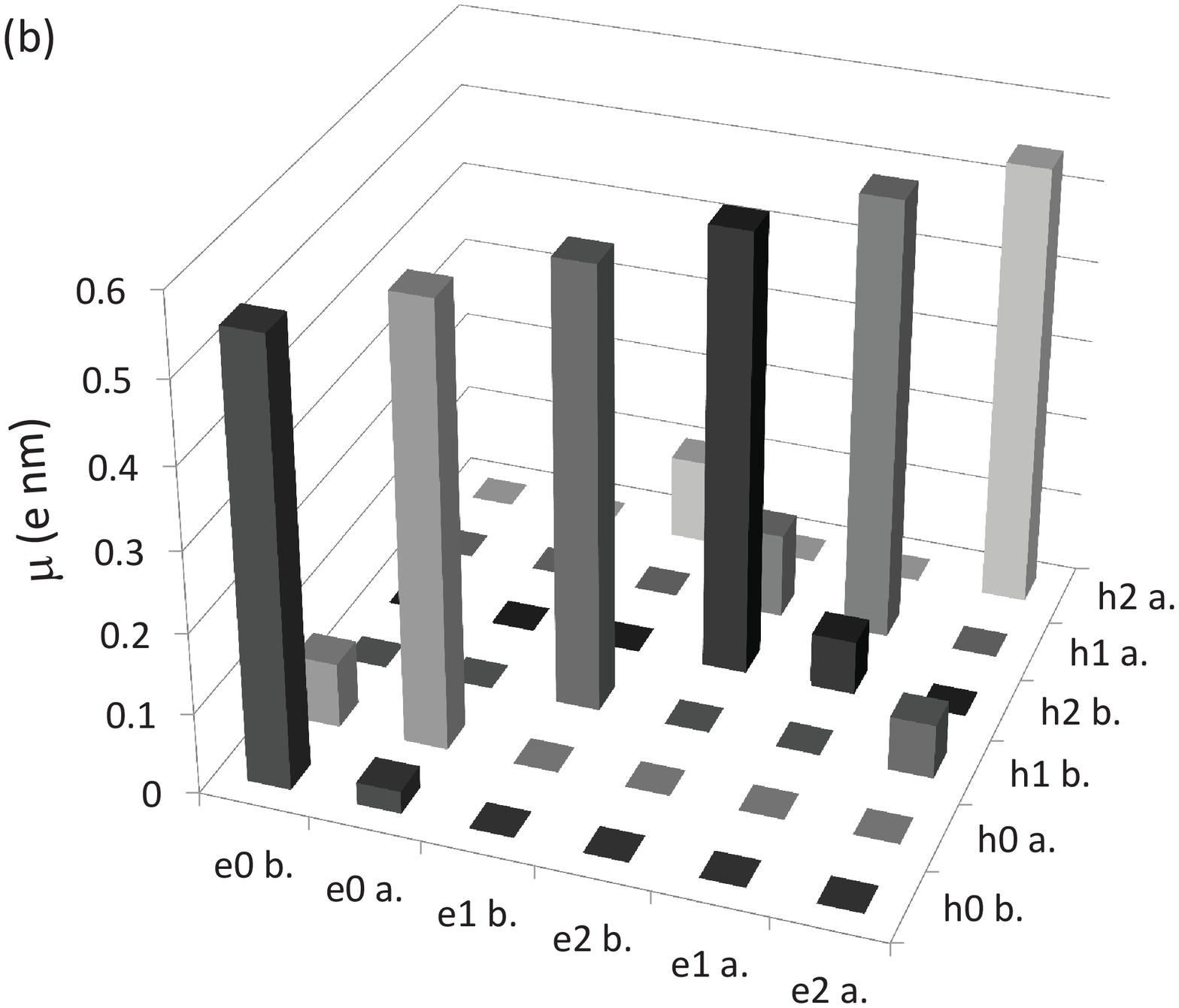}
\includegraphics[trim=1cm 0cm 1cm 0cm,clip,scale=0.3,angle=0]{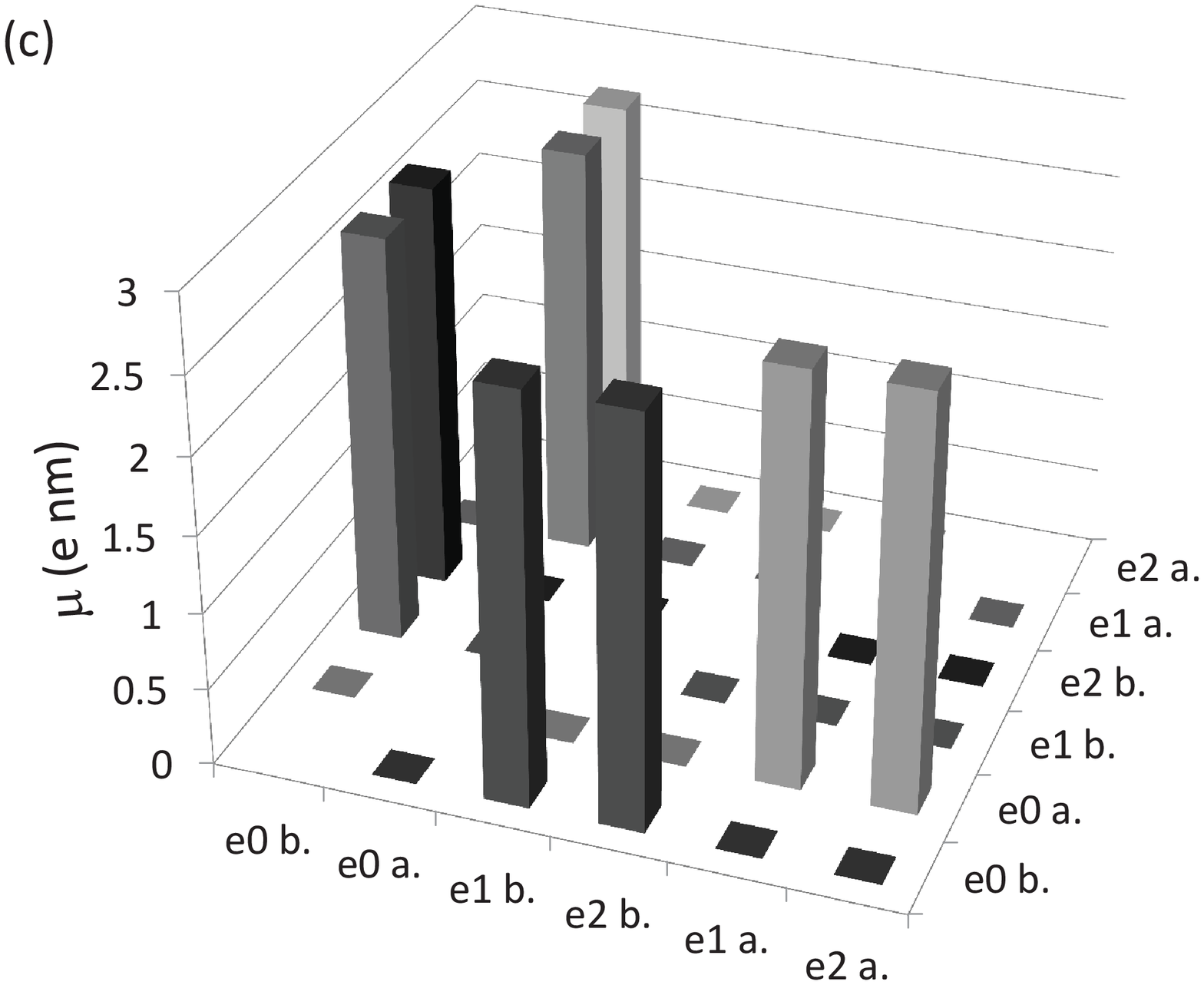}
\includegraphics[trim=1cm 0cm 1cm 0cm,clip,scale=0.3,angle=0]{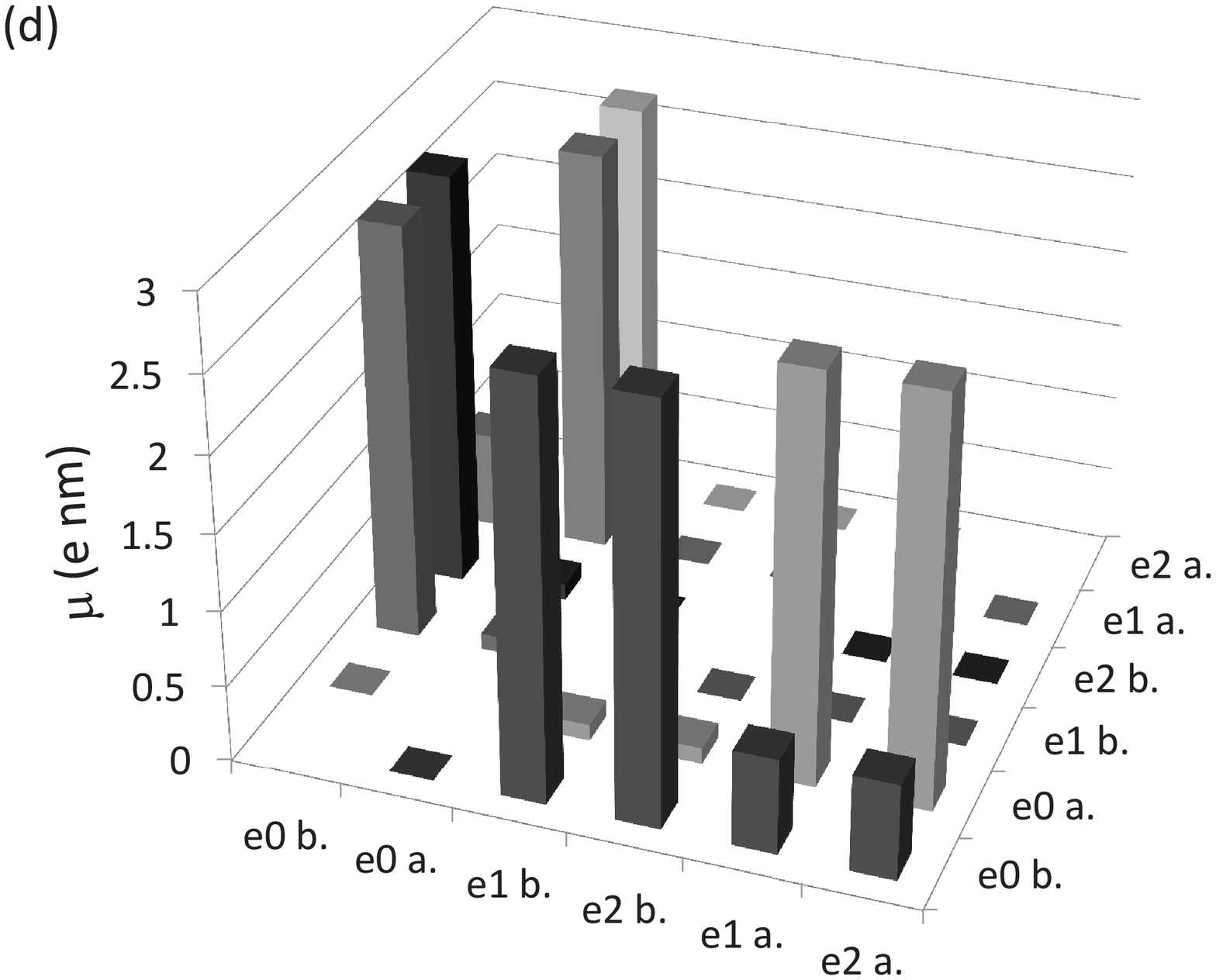}
\caption{Dipole interband (a,b) and intersubband (c,d) matrix elements $\mu$ for QD molecules consisting of two identical (a,c) 
and two different sized (b,d) QDs.
The bonding and antibonding ground states are labeled e0 b. and e0 a. and
the bonding and antibonding degenerate first and second excited states are labeled e1 b., e1 a. and
e2 b., e2 a., respectively. The nomenclature for the holes is similar.}
\label{qdm_dipole1}
\end{figure}

Finally, the dipole interband and the dipole intersubband matrix elements for
the electron states are calculated. For a single QD only diagonal
transitions between electron and hole states for the interband matrix
elements and the ground state to the first or second excited state for the
intersubband matrix elements would be dipole allowed. This behavior carries over to the QD molecule composed of identical QDs, if one regards the bonding
and antibonding states as two separate quantum numbers. For the different sized QDs
the states become more strongly mixed for decreasing QD distance. An
overview of the dipole matrix elements is given in figure \ref{qdm_dipole1}.

\subsection{QD molecule for group-velocity slowdown \label{QDM_asym}}

The last section showed that QD molecules allow one to tailor the electronic structure by the QD molecule design.
The main idea to circumvent the limit of the maximum achievable slowdown in
QDs is to design a QD
molecule that has spatially well
separated electronic wave functions. The spatial separation of the wave functions connected by the quantum coherences minimizes electron phonon and Coulomb matrix elements, which influence the dephasing of this coherence.

First of all, we calculate a small and a large sample QD using $k\cdot p$-theory to adjust
the parameters in the semi-analytical model. For both QDs we assume a
geometry of an obelisk with $\{101\} $ facets. For the small QD
geometry we assume an In$_{0.8}$Ga$_{0.2}$As QD embedded in a GaAs
QW on a wetting layer of thickness 1\,nm. The QD has a
base of $10\,\mathrm{nm} \times 10\,\mathrm{nm}$ and a height of 2\,nm. For this configuration only
the electron and hole ground states are confined. For the large QD
configuration we assume an In$_{0.9}$Ga$_{0.1}$As QD embedded in a
GaAs QW on a wetting layer of thickness $1$~nm. The QD has a base
of $12\,\mathrm{nm}\times 12\,\mathrm{nm}$ and a height of 3\,nm. For this configuration three
electron and three hole states are confined. 
Using the sample QDs we check the parameters in our semi-analytical approach
and calculate the electronic structure of the small and the large QD 
as an intermediary result. 

\begin{figure}[]
\centering
\includegraphics[trim=2cm 0cm 2cm 0cm,clip,scale=0.3,angle=0]{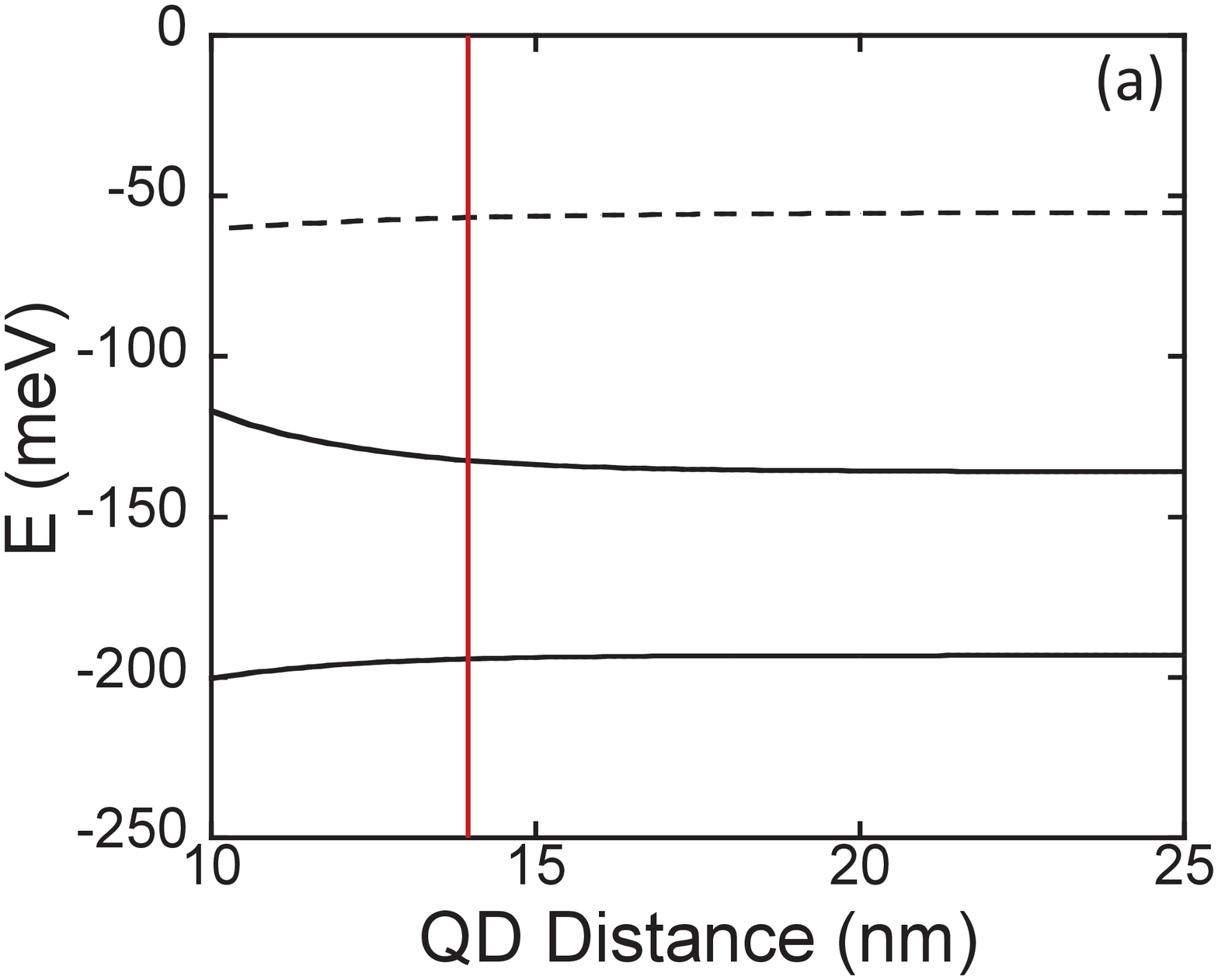}
\includegraphics[trim=2cm 0cm 2cm 0cm,clip,scale=0.3,angle=0]{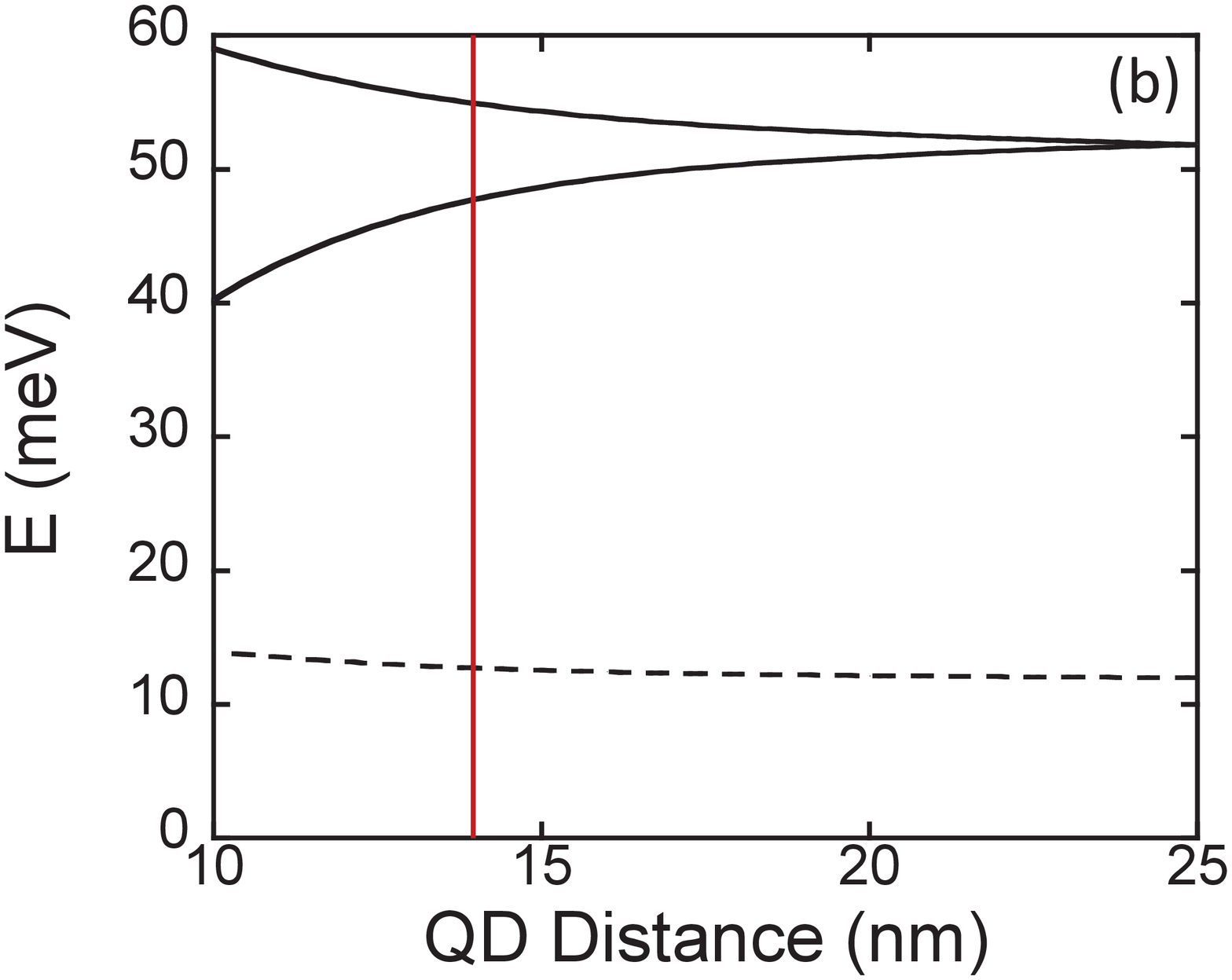}
\caption{Combined conduction (a) and valence band states (b) plotted
  over the distance for the asymmetric double QD molecule described in the text.
The solid lines are the bonding and antibonding ground states and the dashed lines are the
degenerate first and second excited states.}
\label{fig1617}
\end{figure}

\begin{figure}[]
\centering
\includegraphics[trim=5cm 2cm 5cm 4cm,clip,scale=0.6,angle=0]{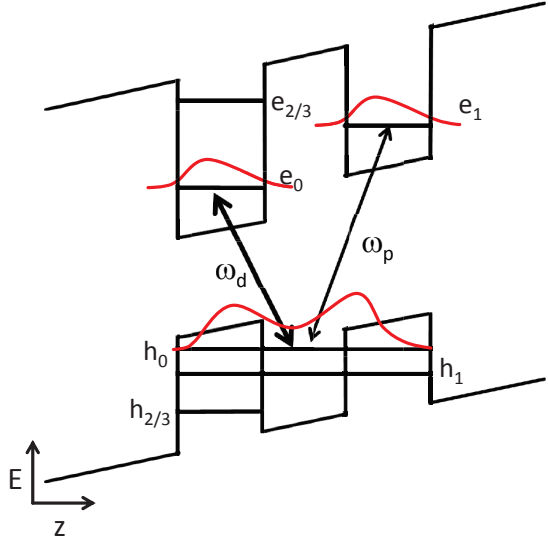} 
\caption{Schematic picture of the lowest-level wave functions and the geometry
  of the asymmetric double QD molecule described in the text.
The resonant probe and drive fields in a $V$-type quantum coherence scheme are also shown.}
\label{fig15}
\end{figure}

\begin{table}
\center
\begin{tabular}{||c||l||}
\hline\hline
\begin{tabular}{cc}
& E$_{e}$ (meV) \\ 
\multicolumn{1}{l}{e$_{0}^{}$} & $-194$ \\ 
\multicolumn{1}{l}{e$_{1}^{}$} & $-132$ \\ 
\multicolumn{1}{l}{e$_{2/3}$} & $-56$%
\end{tabular}
& 
\begin{tabular}{cc}
& E$_{h}$ (meV) \\ 
\multicolumn{1}{l}{h$_{0}^{b}$} & $55$ \\ 
\multicolumn{1}{l}{h$_{1}^{a}$} & $47$ \\ 
\multicolumn{1}{l}{h$_{2/3}$} & $13$%
\end{tabular}
\\ \hline\hline
\end{tabular}
\caption{Electron (e) and hole (h) energies of single-particle states
  in the QD molecule. The bonding and antibonding states formed from hole
  levels of the individual QDs are denoted by h$^{b}$ and h$^{a}$, respectively.}
\label{table-3}
\end{table}

An external electric field in
growth direction opens up the possibility to shift the energy levels of the
QDs. We choose a field that makes the hole
ground states of the small and the large QD degenerate. The other energy
levels are not degenerate due to the different energy spacings of the QDs.
If we couple these two QDs including a suitably chosen static electric field in growth direction, 
we obtain the 
energy eigenvalues for different QD distances depicted in figure~\ref{fig1617}. For large distances between the QDs, we have effectively two separate QDs with different
energy spacing between the states. Other than between the hole ground states, no degeneracy of an energy level
between the small and the large QD occurs. 
An intermediate QD distance allows one to have bonding and antibonding hole
ground states with a sizable energy difference, but without significant mixing with the other states, see figure~\ref{fig1617}.
In particular, the electron ground states of the two
QDs are not significantly mixed. This configuration can be realized with a
dot distance of 14\,nm and a static electric field in growth direction with $E_{\perp }=1.5$\,mV/nm. We have done test calculations for QD molecules which qualitatively confirm the results of our semi-analytical approach. Only the dependence of energy spacing and QD distance is changed somewhat. For instance, a $k \cdot p$-calculation, which includes strain and piezoelectric
effects, yields a result of approximately 10\,nm for this configuration.\cite{dissertation}
The QD molecule with intermediate QD distance is placed in the center of a QW 
with a thickness of 30\,nm. For the QD molecule we obtain four
confined hole and electron states. The energy eigenvalues are 
depicted in figure~\ref{fig15} and compiled in the table~\ref{table-3}.

As already mentioned we have a bonding h$_{0}^{b}$ and antibonding h$_{1}^{a}$ hole ground state and electron ground states e$_{0/1}$
without a significant mixing between the two QDs of the molecule.
Therefore the overlap of the wave functions between the two electron ground
states is small. The transitions between the bonding hole ground state h$_{0}^{b}$ and the electron ground states e$_{0}$ and e$_{1}$ are
dipole allowed with dipole moments of $0.5e$\,nm and $0.2e$\,nm,
respectively. Thus we can realize a $V$-configuration for the QD
molecule as shown in figure~\ref{fig15}. The cw drive is chosen
resonant with the transition of the bonding hole ground state h$_{0}^{b}$ 
and the electron ground state e$_{0}$; the cw probe is chosen resonant to the
transition between the bonding hole ground state h$_{0}^{b}$ and the
electron ground state e$_{1}$. The e$_{0}\leftrightarrow $e$_{1}$ transition
is the quantum coherence transition of the $V$-scheme. 

\section{QD molecules for slowing down light\label{SL_II_3}}

\begin{figure}[]
\centering
\includegraphics[trim=1cm 3cm 1.5cm 5cm,clip,scale=0.6,angle=0]{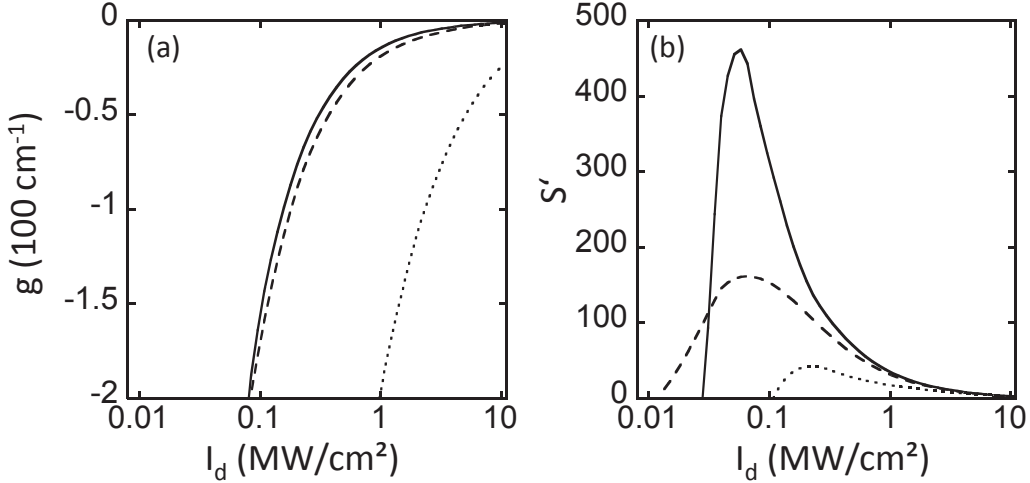}
\caption{Peak gain (a) and peak slowdown (b) versus drive intensity
  for the QD molecule (solid and dashed line) and the deep single QD (dotted
  line). The lattice temperature is 300~K (dashed line) and 150~K (solid
  and dotted line).}
\label{fig19}
\end{figure}

We use a QD molecule $V$-scheme as introduced in section~\ref{QDM_asym} for
slowing down light and compare the results with the results from the deep single
QD.  The peak gain and the peak slowdown are calculated as in section~\ref{SL_II} and, as for the single QD, we plot these quantities in figure \ref{fig19} for different lattice temperatures. 
As in Section \ref{SL_II}, we obtain generally a more efficient slowdown  for
lower temperatures.  The most important result of  figure~\ref{fig19} is the comparison of the deep single QD and the QD
molecule for a lattice temperature of $150$\,K. For similar peak absorption
values an improvement of the peak slowdown by an order of magnitude for the QD
molecule is achieved. This is because in the QD molecule, as compared to the single QD,
the dephasing rate of the quantum coherence is much more reduced than the
dephasing rate of the interband probe polarization. This reduction comes
particularly from the negligible wave-function overlap between the states of the
$\mathrm{e}_{0}\leftrightarrow \mathrm{e}_{1}$ transition. The negligible overlap has a
huge influence on the electron-phonon \emph{and electron-electron} dephasing
contributions, especially for the quantum coherence. This result demonstrates that suitable QD \emph{molecules} may be extremely effective for engineering long dephasing times in self-organized semiconductor QD systems. Additional numerical results on the behavior of the group-velocity slowdown are contained in Ref.~\onlinecite{qdm_prb}.

\section{Conclusion}
We showed, using a microscopic calculation of scattering and dephasing contributions for the coherences involved, that
for group-velocity slowdown in a single QD, a $V$-type scheme  is preferable to a $\Lambda $-type scheme. Here, a deep single QD at low temperatures gives the best results.
We discussed how a simple model for QDs, which is calibrated by numerical calculations, can be used to analyze the electronic properties of  QD \emph{molecules}. In particular,  the electronic structure of QD molecules can be designed to lead to a long lived quantum coherence by effectively separating the electronic states of the $V$ system in different QDs while leading to a delocalized bonding hole state. This design minimizes the dephasing of the quantum coherence between the electronic states and leads to a pronounced increase in group-velocity reduction compared to a single QD.

\begin{acknowledgments}
This work was supported in part by Sandia's LDRD program and Energy Frontier Research Center (EFRC) for Solid-State Lighting Science, funded by U.S. Department of Energy, Office of Science, Office of Basic Energy Sciences. WWC thanks the hospitality of the Technical University Berlin and travel support provided by SFB787.
\end{acknowledgments}

\end{document}